\documentclass[aps,pre,twocolumn,superscriptaddress,floatfix]{revtex4}

\usepackage{graphicx,latexsym}
\usepackage{xcolor}

\begin{document}


\title{The rolling and slipping of droplets on superhydrophobic surfaces}


\author{A. F. W. Smith}
\affiliation{MacDiarmid Institute for Advanced Materials and Nanotechnology, Department of Physics, University of Auckland, Auckland 1142, New Zealand}
\author{K. Mahelona}
\affiliation{Wai Foundation, Taipa 0483, New Zealand}
\author{S. C. Hendy}
\affiliation{MacDiarmid Institute for Advanced Materials and Nanotechnology, Department of Physics, University of Auckland, Auckland 1142, New Zealand}
\affiliation{Te P\={u}naha Matatini,  Department of Physics, University of Auckland, Auckland 1142, New Zealand}



\date{\today}

\begin{abstract}
The leaves of many plants are superhydrophobic, a property that may have evolved to clean the leaves by encouraging water droplets to bead up and roll off. Superhydrophobic surfaces can also exhibit reduced friction and liquids flowing over such surfaces have been found to slip in apparent violations of the classical no-slip boundary condition. Here we introduce slip into a model for rolling droplets on superhydrophobic surfaces and investigate under what conditions slip might be important for the steady state motion. In particular, we examine three limiting cases where dissipation in the rolling droplet is dominated by viscous dissipation, surface friction, or contact line friction. We find that in molecular dynamics simulations of droplets on ideal superhydrophobic surfaces with large effective slip lengths, contact line dissipation dominates droplet motion. However, on real leaves, droplet motion is likely to be dominated by viscous shear, and slip, for the most part, can be neglected.
\end{abstract}


\maketitle

%
\section{Introduction}
Superhydrophobic surfaces that exhibit the Lotus effect \cite{Barthlott97} are of interest both for their role in biology \cite{Barthlott97b} and for their potential technological applications \cite{Quere03}. Superhydrophobic surfaces, including the surfaces of many plant leaves \cite{fritsch2013superhydrophobic}, combine nano- and micro-scale roughness with a hydrophobic surface coating to achieve contact angles of up to 160$^{\circ}$. The Lotus effect is thought to benefit plants by helping to keep leaves clean; droplets of moisture bead up and eventually roll down leaves, entraining dirt and contaminants as they go \cite{Quere05}. Indeed, experiments have found that the fact that droplets roll rather than slide down superhydrophobic surfaces \cite{Quere99} makes them more likely to remove contamination along the way.   

More recently, however, flows over superhydrophobic surfaces have been studied because they effectively violate the classical no-slip boundary condition \cite{Cecile03}. When in the Cassie state, droplets or larger scale flows are lubricated by an entrapped layer of air, leading to large effective slip lengths with drag only occurring at the few points of the surface where the flow makes contact with the substrate \cite{Rothstein2010a}. On such superhydrophobic surfaces effective slip lengths of tenths to tens of microns have been observed \cite{Choi06,Cecile06}, scaling proportionally to the typical microstructural length scale \cite{Cecile06,Ybert07}. In some experiments, on highly ideal superhydrophobic surfaces, slip lengths of hundreds of microns have been measured \cite{Kim08}.

The canonical model for the steady state motion of droplets on superhydrophobic surfaces, proposed by Mahadevan and Pomeau \cite{Mahadevan99}, assumes a purely rolling motion and results in an expression for the droplet velocity that arises from viscous dissipation due to shear in the vicinity of the area of contact between the droplet and the surface. In sufficiently small droplets, however, shear stress at the surface of contact may also generate slip. In this paper, we investigate the conditions under which droplets may slip as they move on superhydrophobic surfaces, whether leaves or engineered surfaces.

\begin{figure}
\resizebox{0.9\columnwidth}{!}{\includegraphics{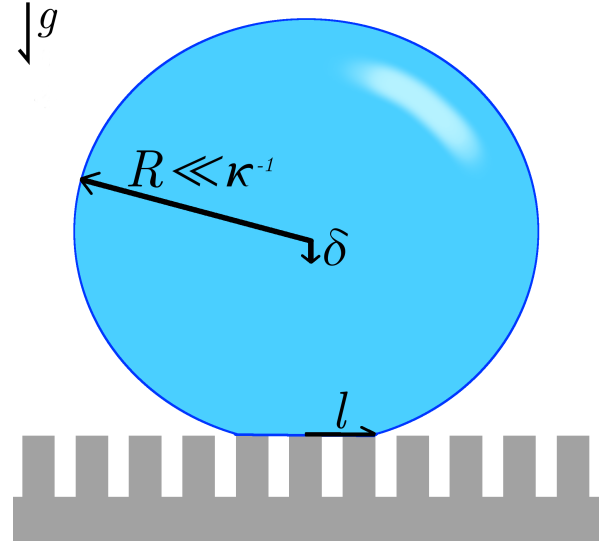}}
\caption{A droplet in the Cassie state on a superhydrophobic surface. Although the droplet radius is below the capillary length $R \ll \kappa^{-1}$, the droplet is depressed by a distance $\delta$ by gravity (gravitational potential energy is converted to surface energy), to form a contact zone with the surface of radius $\ell$.}
\label{Fig1}
\end{figure}

Mahadevan and Pomeau's model considers droplets supported on a superhydrophobic surface that are nearly spherical  (see Figure~\ref{Fig1}). This will only be a reasonable approximation if the droplet radius is smaller than the capillary length, $\kappa^{-1} = (\gamma/\rho g)^{1/2}$, where $\gamma$ is the droplet surface tension (for water $\simeq 2$ mm) and $\rho$ is the fluid density. The radius of the contact zone $\ell$ is determined by a balance in the loss in gravitational potential energy that results from the lowering of the droplet centre of mass with the corresponding creation of extra surface area. In this case the contact zone radius $\ell$ can be shown to be $\sim R^2/\kappa^{-1}$ \cite{Mahadevan99}.

\begin{figure}
\resizebox{\columnwidth}{!}{\includegraphics{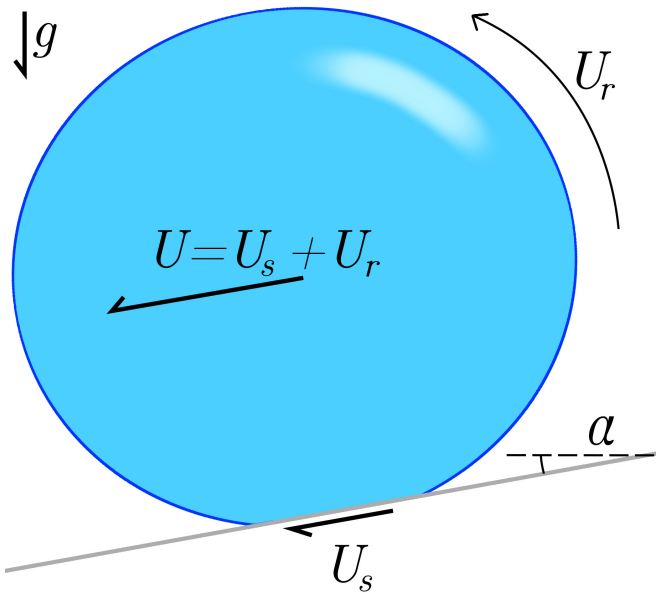}}
\caption{A droplet rolling down a superhydrophobic surface tilted at an angle $\alpha$ to the horizontal with respect to the direction of the body force $m\vec{g}$. The centre of mass velocity of the droplet, $U$, is assumed to be equal to the sum of rolling velocity $U_r$ and any slip velocity $U_s$.}
\label{Fig2}
\end{figure}

The droplets are considered to be moving on a surface tilted at an angle $\alpha$ to the horizontal. The centre-of-mass velocity of the droplet $U$ is equal to its rolling velocity $U_r$ (Figure~\ref{Fig2}), with a no-slip condition resulting in zero slip at the surface ($U_s = 0$). In the contact zone, shear induces a velocity gradient $|\nabla u| \sim U/R$ and generates viscous dissipation. The rate of dissipation due to viscous shear in the contact zone will then be given by 
\begin{equation}
P_v = \int_V{\mu (\nabla u)^2 dV} = \mu \int_V{\left(\frac{U}{R}\right)^2 dV} \sim  \frac{\ell^3 \mu U^2}{R^2}
\end{equation}
where the integral has been taken over the volume ($V\sim \ell^3$) of the contact zone. 

In the steady-state, the rate of viscous dissipation must balance the rate of loss of gravitational potential energy, giving:
\begin{equation}
\rho g R^3 U\sin{(\alpha)} \sim \frac{\ell^3\mu U^2}{R^2}.
\end{equation}
Solving for $U$ yields the steady-state centre-of-mass velocity:
\begin{equation}
\label{MPviscous}
U \sim \frac{\gamma}{\mu} \frac{\kappa^{-1}}{R} \sin{(\alpha)},
\end{equation}
which shows that $U$ scales in inverse proportion to the droplet radius $R$.

However, for small droplets on superhydrophobic surfaces, the no slip boundary condition ($U_s = 0$) may not be a good approximation. Indeed, some experiments have found that on highly engineered superhydrophobic surfaces, rolling can be entirely suppressed \cite{sakai06,sakai09}. Other works have focused on the opposing limit, where droplet motion is constrained by pinning-depinning events or contact line friction as the droplet progresses across the surface. These approaches vary from analysis with molecular kinetic theory \cite{Blake69,Blake02,Blake06} to phenomenological models (where, for example, the effect of pinning-depinning events is captured by a 'sliding resistance', tied to the droplet's dynamic receding contact angle \cite{Olin2013}) to consideration of the energy losses involved in the rupturing of capillary bridges found at the receding edge of the droplet \cite{Butt2017}. 

Droplet motion in the presence of slip has been studied previously by simulation, but these studies have generally either not explicitly considered superhydrophobic structures \cite{Servantie2008, Annapragada2012}, or were restricted to two dimensions \cite{Mognetti2010, Li2013, Wind-Willassen2014}, which affects both droplet dynamics and the spectrum of available surface geometries. A recent study \cite{Karapetsas2016} in three dimensions simulated droplet motion using a continuum approach, but applied an effective slip boundary condition that only indirectly includes the effects of surface microstructure.

In this paper we report on three dimensional molecular dynamics simulations of the motion of droplets on superhydrophobic surfaces using molecular dynamics, and extensions of Mahadevan and Pomeau's model to include slip, while considering the role of surface friction and contact line dissipation. In particular, we have conducted molecular dynamics simulations of droplets moving in steady-state on a superhydrophobic surface in response to a gravitational force. The simulations allow us to identify a regime where the speed of the droplets is proportional to droplet size, rather than the inverse of the droplet size as given by equation~\ref{MPviscous}. We extend this model \cite{Mahadevan99} to incorporate an effective Navier slip boundary condition \cite{Hendy07} and we consider scenarios where the steady-state motion of the droplet is either dominated by friction at the solid-liquid interface or by contact line dissipation. We show that it is most likely the latter mechanism that limits the steady-state motion of the simulated droplets. 

\section{Molecular Dynamics Simulations}
Molecular dynamics simulations of simple liquid droplets on superhydrophobic surfaces were carried out using LAMMPS \cite{LAMMPS}. The liquid and surface were modeled using the Lennard-Jones (L-J) 6-12 potential
\begin{equation}
V(r) = 4\epsilon\left[ \left(\frac{\sigma}{r}\right)^{12} - \left(\frac{\sigma}{r}\right)^6 \right].
\end{equation}
In what follows we work in L-J units. We used Lennard-Jones potentials to model intra-droplet ($\epsilon_{LL} = 1.75$) and droplet-substrate interactions ($0.09375 < \epsilon_{SL} < 0.25$). Given the constant $\epsilon_{LL}$, altering $\epsilon_{SL}$ affected both the observed contact angle and the effective slip length of the surface. We find static contact angles between $\sim140^{\circ}$ for $\epsilon_{SL}=0.25$ and $\sim160^{\circ}$ for $\epsilon_{SL}=0.09375$, consistent with previous results \cite{Yong2009}. Droplets were subject to a small body force, with values between $g=1\times10^{-4}\epsilon\sigma^{-2}\tau^{2}$ and $8\times10^{-3}\epsilon\sigma^{-2}\tau^{2}$ selected and applied at an angle of $10^\circ$ from them vertical giving a tilt angle of $\alpha = 10^\circ$.

\begin{figure}
\resizebox{\columnwidth}{!}{\includegraphics{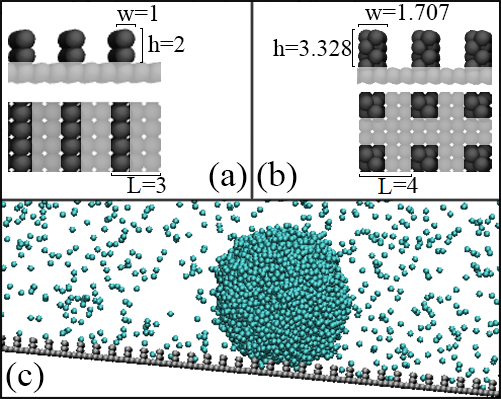}}
\caption{The height h, width w, and period L of the surface features for the ridged (a) and pillared (b) surfaces are presented above, with all values given in terms of the lattice spacing $\sigma$. The particle diameter is shown as $1\sigma$, matching the lattice spacing. (c) shows a snapshot of a simulated droplet ($R=10\sigma$) rolling in the Cassie state on a ridged superhydrophobic surface. Although the droplet radius is below the capillary length $R \ll \kappa^{-1}$, the droplet is compressed a distance $\delta$ by gravity (gravitational potential energy is converted to surface energy), to form a contact zone with the surface of radius $\ell$, spanning multiple surface features.}
\label{FigMD}
\end{figure}

We used both ridged and pillared surfaces in our simulations (see Figure~\ref{FigMD}). The ridged surface consisted of straight, raised ridges with period $L$ arranged perpendicular to the surface normal and the direction of the body force. The pillared surface consisted of raised posts, arranged in a square lattice with period $L$. The atoms comprising the posts themselves were arranged in four layers of two particles, with the atoms occupying opposing corners of a cube with side-length $\frac{1}{\sqrt{2}}$ in a tetrahedral fashion. The exact dimensions are detailed in table \ref{surf_properties} and shown in panels (a) and (b) of figure \ref{FigMD}.

\begin{table}[htp]
\begin{center}
\begin{tabular}{|l|c|c|c|c|}
\hline
 & Height, h & Width, w & Period, L & Area Fraction, $\phi$\\
\hline 
Ridges & 2$\sigma$ & 1$\sigma$ & 3$\sigma$ & 0.33\\
\hline
Posts & 3.328$\sigma$ &  1.707$\sigma$&  4$\sigma$ & 0.18\\
\hline
\end{tabular}
\end{center}
\caption{The dimensions of the ridged and pillared surfaces used (see Figure~\ref{FigMD}), approximating the effective surface particle diameter as $1\sigma$. The  period $L$ of the pattern was $3\sigma$ for the ridged surface and $4\sigma$ for the pillars.}
\label{surf_properties}
\end{table}%

The temperature was controlled by a Langevin thermostat \cite{Schneider78} applied to all fluid particles in the direction perpendicular to the surface normal and the direction of the body force with a set temperature of $1.1 \frac{\epsilon}{k_B}$. This temperature was chosen to ensure that the droplet was always completely liquid for the range of droplet sizes investigated. Vapor was present in equilibrium with the rolling droplets.

There are several ways one can calculate the viscosity in molecular dynamics \cite{Hess:2002p8998}. We used a Green-Kubo approach, in tandem with simulated Poiseuille flows of the confined fluid. The calculated bulk viscosity $\mu$ was 3.5$\sigma^{-1}\tau^{-1}$, similar to those reported in other works \cite{Rowley:1997p8980,Tankeshwar:1988p8991}. Both calculations were made for a bulk system at higher temperatures than the droplets, as the droplets in our simulations are super-cooled relative to the bulk freezing temperature, remaining liquid due to their finite size. 

While we anticipate that the viscosity within our droplets will be slightly different to these calculations, our interest is in the observed scaling of droplet velocities with their radii, and so the estimates here are adequate for our purposes. The surface tension was found to be $\sim$0.6 $\epsilon \sigma^{-2}$ by examining the relationship between drop size to its internal energy \cite{Lee:1987p8281}. Again, our estimates are of the same order as those reported elsewhere \cite{Potoff:2000p8106}. The particle density $\rho$ within the droplet was around 1.1$\sigma^{-3}$ for all droplet sizes discussed here. Based on these estimates of the liquid properties, we expect the quasi-spherical approximation to hold for the droplet's shape, as the Weber and Capillary numbers ({\em We} and {\em Ca}) are less than unity in the simulations here. 

The motion of the droplets were simulated until a steady state velocity was reached and then runs were extended to allow for data collection. Depending on the body force $g$ applied, this would take between $10^4$ and $10^6$ time steps. The Lennard-Jones particles in the droplet (as opposed to those in the vapor - see Fig.~\ref{FigMD}) were identified in post-processing using a distance-threshold approach, with all droplet mass, density, radius and motion statistics being calculated from the distribution of positions and velocities belonging to the associated particles. The rolling velocity $U_r$ was calculated by fitting a straight line to the upper half of the droplet velocity profiles to obtain the shear rate $\nabla{u}$ and rolling velocity $U_r = R \, \nabla{u}$. The slip velocity was then calculated as $U_s = U-U_r$, where $U$ is the center-of-mass velocity of the droplet.

\section{Results}
\begin{figure}
\resizebox{\columnwidth}{!}{\includegraphics{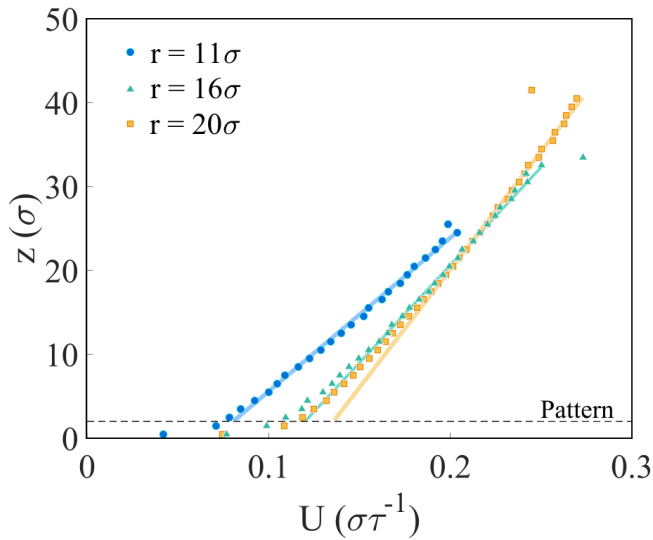}}
\caption{The figure shows the velocity profiles of droplets with initial radii of 11$\sigma$ (blue circles), 16$\sigma$ (green triangles), and 20$\sigma$ (red squares) in steady-state motion on a ridged surface with $g=1\times 10^{-3}\sigma\tau^{-2}$ and $\epsilon_{SL}=0.125\epsilon$. Straight lines have been fitted to the top half of each profile (the good fit indicates rolling) and projected down towards the surface for reference. A non-zero slip velocity $U_{slip}$ can be seen at the pattern's surface (indicated by the dashed line at $z=2\sigma$) for each of the droplet sizes, with the deviation between the actual slip velocity and the linearly-fitted lines increasing as the droplet radius increases. Error bars are determined from the standard deviation, with $U$ values taken from 5000 simulation frames, and $U_s$ and $U_r$ constructed from 40 hundred-frame bins.}
\label{FigVelPro}
\end{figure}

Figure~\ref{FigVelPro} shows the steady-state velocity profiles for three drops with differing radii ($R=$ 11 $\sigma$, 16$\sigma$ and 20$\sigma$) simulated with $\epsilon_{SL}=0.125\epsilon$ (unless otherwise stated, this value of $\epsilon_{SL}$ is used throughout) on a ridged substrate tilted at an angle $\alpha = 10^\circ$ with $g=1\times 10^{-3}\sigma\tau^{-2}$. The average velocity of atoms within a $1\sigma$-thick slice are shown as a function of height $z$ above the base of a ridged substrate. Note that the velocity profile at and just below the top of each droplet is linear, indicating rolling motion, while slip is evident at the substrate. The larger drops have a higher steady state velocity than the smaller drops, with $U_s$ (the velocity at the top of the patterned surface) increasing as droplet size increases while $U_r$ (which is proportional to the slope of the linear profile) remain relatively constant with size.

\begin{figure}
\resizebox{\columnwidth}{!}{\includegraphics{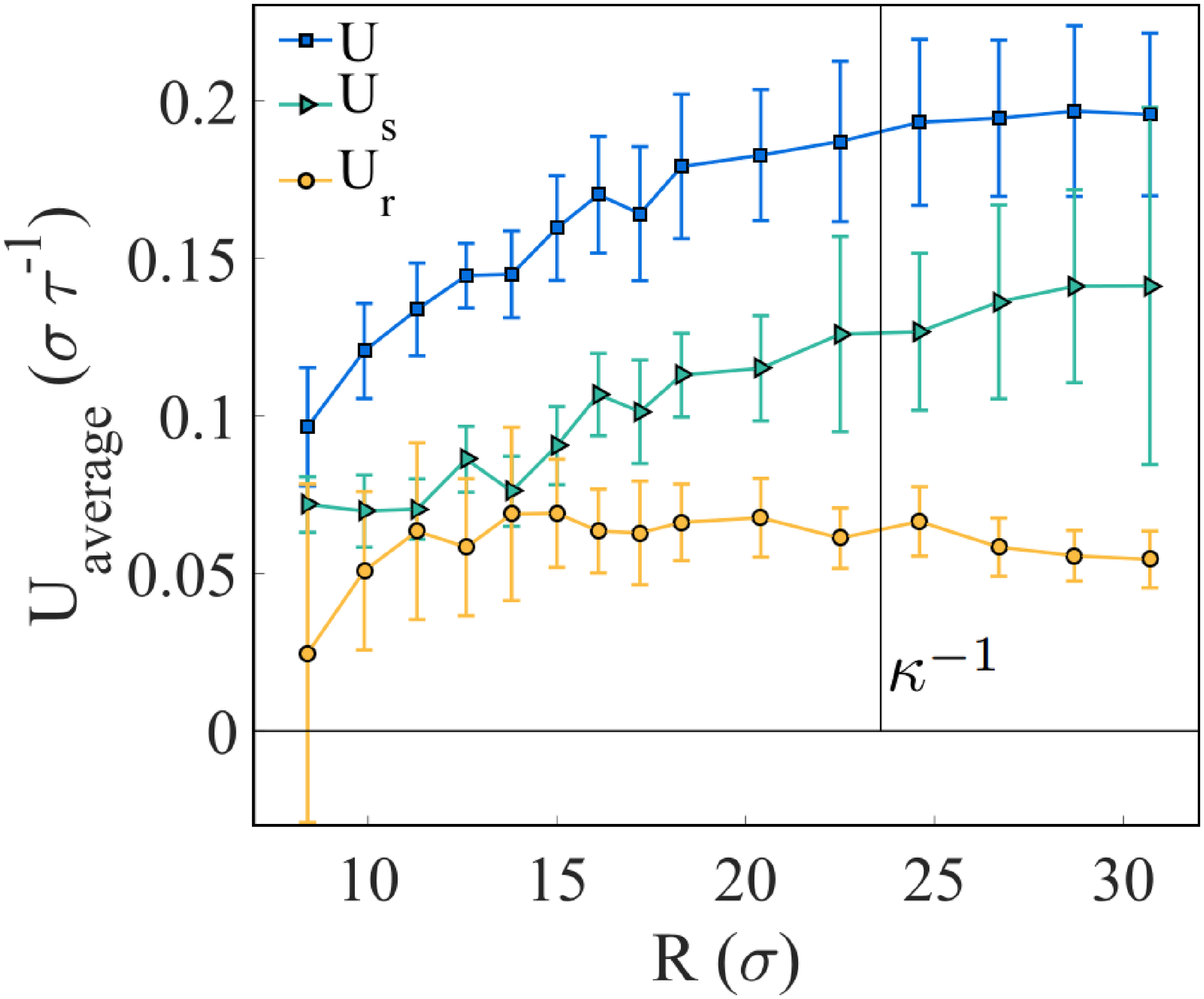}}
\caption{Plot of time-averaged droplet center-of-mass velocities $U$ versus the various droplet radii on a ridged surface with $g=10^{-3}\sigma\tau^{-2}$ and $\epsilon_{SL}=0.125\epsilon$. Droplets larger than the radius indicated by the right-hand vertical line are larger than the capillary length $\kappa^{-1}$ and so can be expected to deform and depart from spherical shape assumed by the model. Error bars indicate the standard deviation associated with the simulated data.}
\label{FigVelR}
\end{figure}

Figure~\ref{FigVelR} shows $U$ (along with $U_s$ and $U_r$) for a wider size range of droplets on ridged surfaces ($R = 9\sigma$ to $31\sigma$). We note a steady increase in center-of-mass velocity with droplet size up to about $R \sim 20 \sigma$. Above this the velocity plateaus as the droplet radius approaches the capillary length. At this point the droplet starts to deviate from its spherical shape, and both $Re$ and $We$ begin to approach unity. Nonetheless, the increase in velocity with droplet size below the capillary length is very different to the behavior predicted by equation~\ref{MPviscous}. We also note that it is largely the slip velocity $U_s$ that increases with droplet size. Simulations performed on the pillared surfaces exhibit similar linear scaling with $R$, although the centre-of-mass velocities are somewhat higher (we expect pillared surfaces to exhibit higher effective slip lengths as the area fraction $phi$ covered by the pillars is less than the ridges; table ~\ref{surf_properties}). 

\begin{figure}[htb]
\center
\resizebox{\columnwidth}{!}{\includegraphics{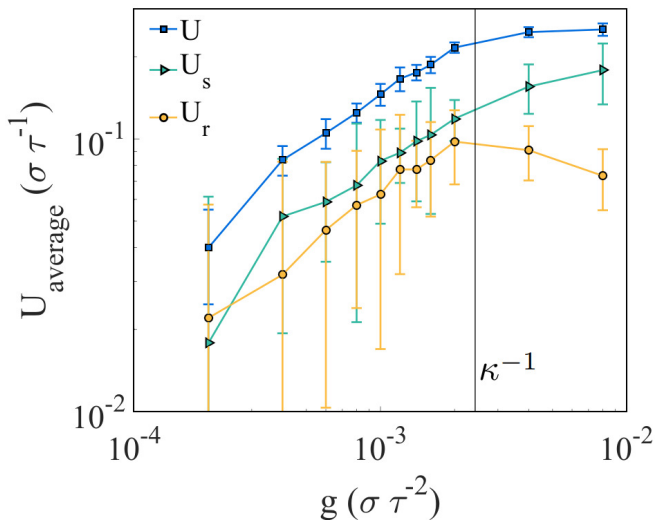}}
{\caption{A log-log plot of the steady-state center-of-mass velocities for droplets with radii fixed at 16$\sigma$ and $\epsilon_{SL}=0.125 \epsilon$, as $g$ is increased from $1\times10^{-4}$ to $10^{-2}\sigma\tau^{-2}$. The simulations were performed on the pillared surface. The apparent linear region between $\sim 4\times10^{-4}\sigma\tau^{-2}$ and $2\times10^{-3}\sigma\tau^{-2}$ corresponds to a power law $U\propto g^\beta$ with an exponent $\beta\approx 0.59$.}
\label{FigVelg}}
\end{figure}

Figure~\ref{FigVelg} shows $U$ for a 16$\sigma$ radius droplet on a {\em pillared} surface as the magnitude of body force $g$ is changed. For small $g$ ($\leq 2 \times 10^{-3} \sigma \tau^{-2}$) we see that the simulated velocity scales close to $\sqrt{g}$ (a power law fit finds an exponent 0.59). Above this threshold the dependence on $g$ is much weaker. We note that the magnitude of $g$ affects the capillary length: given the droplet radius of $R=16\sigma$, we expect the droplet to lose its spherical shape for values of $g$ exceeding $\frac{\gamma}{\rho R^2} = 2\times10^{-3}\sigma\tau^{-2}$. This transition is reflected in the data's loss of linearity for the two highest values of $g$ that were simulated, $4$ and $8\times 10^{-3}\sigma\tau^{-2}$. 

\begin{figure}[htb]
\center
\resizebox{\columnwidth}{!}{\includegraphics{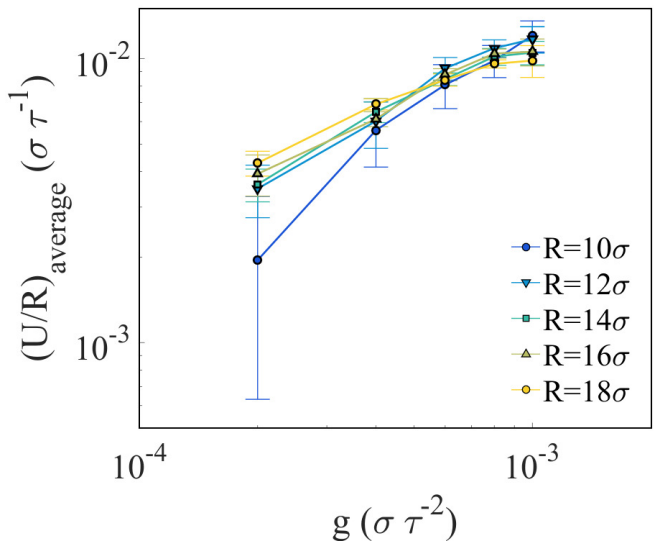}}
{\caption{The steady-state center-of-mass droplet velocities scaled by radius for radii between 10$\sigma$ and 18$\sigma$ on a ridged surface with $\epsilon_{SL}=0.125 \epsilon$, as $g$ is increased from $2\times10^{-4}$ to $1\times10^{-3}\sigma\tau^{-2}$. The log-log plot confirms the $R/\kappa^{-1}$ scaling of the centre-of-mass velocity for droplets smaller than the capillary length.}
\label{FigVelgscaling}}
\end{figure}

Figure~\ref{FigVelgscaling} demonstrates similar scaling behaviour for droplets on the ridged surface. In this figure we have plotted $U/R$ versus $g$ on log-log axes to demonstrate that the $U \sim R/\kappa^{-1}$ scaling holds for droplets with radius $R = 10-18 \sigma$ (albeit noting the one outlier for the $R=10 \sigma$ droplet under very weak body force, where the observed that the droplet has a tendency to bounce). We note that this is the inverse of the scaling behaviour predicted by Eqn.~\ref{MPviscous}.

\begin{figure}[htb]
\center
\resizebox{\columnwidth}{!}{\includegraphics{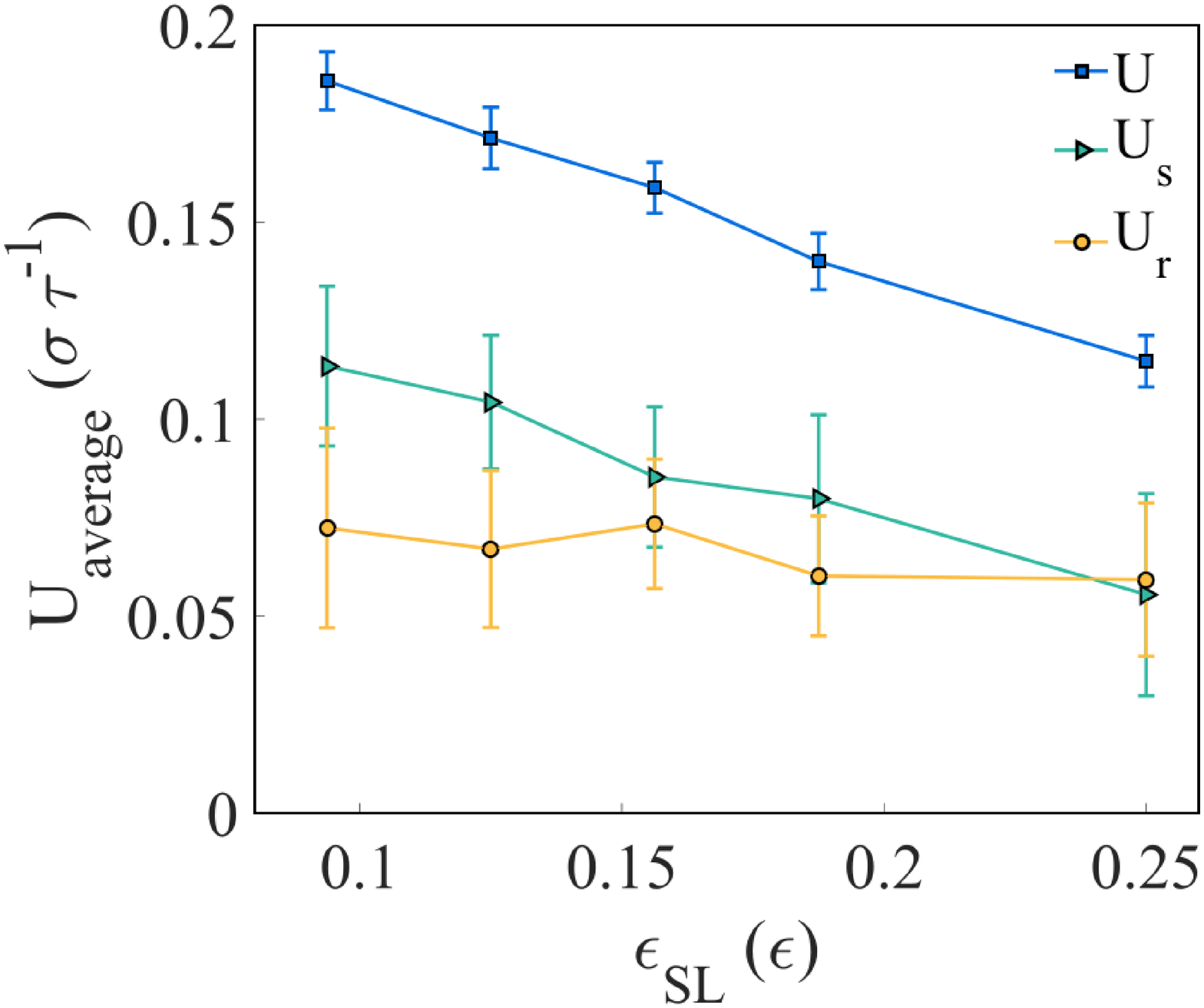}}
{\caption{$U$ and $U_{s}$ are found to decrease as the strength of the liquid interaction with the ridged surface, characterised by $\epsilon_{SL}$, grows larger. Here, $g$ is $10^{-3} \sigma\tau^{-2}$ and $R$ is $16 \sigma$.}
\label{FigEps}}
\end{figure}

For the majority of results presented above we used a value of $\epsilon_{SL} = 0.125 \epsilon$, but figure~\ref{FigEps} shows how $U$ varies as $\epsilon_{SL}$ is varied between $0.09375 \epsilon$ and $0.25 \epsilon$ for the ridged surface. We also note a steady decrease in $U$ and $U_s$, while $U_r$ remains approximately constant. Simulations on the equivalent pillared surfaces show similar behaviour. Equation~\ref{MPviscous} does not make predictions about how $U$ should change with solid-liquid interaction strength, but as $\epsilon_{SL}$ increases, of course, the approximations that lead to this equation (as well as aspects of the theory developed below) would be expected to lose validity. 

\section{Extended Models}
The molecular dynamics simulations in the previous section show that the droplet is slipping in the contact region, something that is not taken into account in equation~\ref{MPviscous}. Our starting point in this section is to introduce slip into the model Mahadevan and Pomeau\cite{Mahadevan99} via an effective Navier slip boundary condition (assumed to hold over the surface of the contact region). This relates the slip velocity $U_s$ at the surface to the velocity gradient near the contact zone: $U_s = b \left| \nabla u \right|_s$, so that $U_s = U_r b/R$ and $U \sim U_r (1+b/R)$. If $R \gg b$ then $U_s/U_r \rightarrow 0$ and we recover the model of of Mahadevan and Pomeau.

Here the slip length $b$ should be considered an effective slip length \cite{Ybert07, Hendy07} i.e. the slip length of some homogeneous surface that would result in the same frictional shear as the inhomogeneous superhydrophobic surface when averaged over a sufficiently large area. For our purposes, the use of an effective slip length is justified provided the size of the contact region, $\ell$, is much larger than the structuring of the superhydrophobic surface. For a surface composed of arrays of posts, with post spacing $L$, the effective slip length $b$ is expected to scale as $L/\phi^{1/2}$ where $\phi$ is the area fraction of the surface covered by the posts \cite{Ybert07}. In this case we are justified in using an effective slip length provided $\ell \gg L \sim \phi^{1/2} b$. 

We begin by writing the centre of mass velocity $U$ as the sum of the rolling velocity and the slip velocity: $U=U_r+U_s$. The viscous dissipation is then given by
\begin{equation}
P_v = \int_V{\mu (\nabla u)^2 dV} \sim \ell^3 \mu \left(\frac{U_r}{R}\right)^2,
\end{equation}
which is balanced by the rate of loss of gravitational potential energy in the steady state: 
\begin{equation}
\rho g \sin{(\alpha)} \, UR^3  \sim \frac{\ell^3\mu}{R^2}U^2\left(1+\frac{b}{R}\right)^{-2}.
\end{equation}
This leads to the following expression for the steady-state velocity $U$:
\begin{equation}
U \sim \frac{\gamma}{\mu} \frac{\kappa^{-1}}{R} \left(1+b/R\right)^2 \sin{(\alpha)}.
\end{equation}
Note that the droplet velocity is inversely proportional to the droplet radius $R$ (as in \ref{MPviscous}), with a correction due to slip that scales as $b/R$. In the limit of small slip, $b \ll R$, then the expression for the velocity $U$ reduces to that of Mahadevan and Pomeau \cite{Mahadevan99} plus corrections of order $b/R$: 
\begin{equation}
U \sim \frac{\gamma}{\mu} \frac{\kappa^{-1}}{R} \sin{(\alpha)} + O(b/R).
\label{viscous}
\end{equation}
In the large slip limit, $b \gg R$, the expression suggests that $U \sim \frac{\gamma}{\mu} \frac{\kappa^{-1}b^2}{R^3} \sin{(\alpha)}$, although we will show below that in this limit viscous dissipation will not be the dominant mechanism. Regardless, neither of these expressions describe the increase in steady-state velocity with $R$ and $g$ that was seen in the simulations from the previous section (Fig.~\ref{FigVelgscaling}).  

We now consider two other potential sources of dissipation: interfacial friction, associated with fluid sliding over the surface, and contact-line friction, due to contact-line pinning-depinning at surface features. The rate of dissipation due to interfacial friction over the contact zone surface can be expressed as
\begin{equation}
P_f = \int_A{\sigma_s \, u \, dA}
\end{equation}
where $\sigma_s$ is the shear stress at the contact zone surface, and the integral is taken over this surface. The shear stress at the contact surface is given by the Navier slip-boundary condition: $\sigma_s = \mu \left| \nabla u \right| = (\mu / b) U_s$. Thus we estimate  
\begin{equation}
P_f = \frac{\mu}{b} \int_A{U_s^2 dA} \sim \ell^2 \frac{\mu}{b} U_s^2
\end{equation}
Along the contact line, the dissipation can be written as \cite{deGennes92} 
\begin{equation}
P_c = \int_c{\zeta U_s^2 dl} \sim \ell \zeta U_s^2
\end{equation}
where $\zeta$ is an effective contact-line friction coefficient and the integral is taken over the contact line perimeter. 

The ratio of the rates of interfacial frictional dissipation to viscous dissipation is given by
\begin{equation}
\frac{P_f}{P_v} \sim \frac{b}{\ell} \sim \frac{b\kappa^{-1}}{R^2} 
\end{equation}
so when $\ell \gg b$, the frictional term will become unimportant. Thus viscous dissipation can dominate only when $b \ll \ell$ or $b \ll \ell \sim R^2/\kappa^{-1} \ll R$. 

Similarly, the ratio of the rate of contact line dissipation to that due to interfacial friction is given by
\begin{equation}
\frac{P_c}{P_f} \sim  \frac{\zeta}{\mu} \frac{b}{\ell}. 
\end{equation}
Thus, the importance of contact line dissipation depends on the ratio of $\zeta / \mu$, in addition to the droplet size $R$ and slip length $b$. We would expect contact line dissipation to dominate when both $b \gg \ell$ and $\zeta / \mu \gtrsim 1$.  

We now examine two limiting cases for steady-state droplet motion, where either frictional or contact-line dissipation dominates the balance with gravitational potential energy loss. In the first case, which will arise when $\mu/\zeta \gg b/\ell \gg 1$, frictional dissipation will dominate, leading to the following balance: 
\begin{equation}
\rho g \sin{(\alpha)} \, UR^3 \sim \ell^2 \frac{\mu}{b} U^2\left(1+\frac{R}{b}\right)^{-2}.
\end{equation}
This gives
\begin{equation}
U \sim \frac{\gamma}{\mu} \frac{R}{b}\left(1+b/R\right)^2 \sin{(\alpha)}.
\label{friction}
\end{equation}
Here we see that the velocity is proportional to the droplet radius for small slip $b \ll R$ but inversely proportional to radius for large slip $b \gg R$. (We note that use of an effective slip length in this large slip limit is justified for a surface with posts provided $b \sim L/\phi^{1/2} \gg \ell \gg L$). Interestingly, the steady-state velocity is independent of the capillary length so does not scale with $g$. This is somewhat counter-intuitive (and it is not consistent with what is observed in the simulations as is evident from figure~\ref{FigVelgscaling}), but arises because the frictional losses scale with $g$ in the same way as the loss of gravitational potential energy. 

The remaining limit is one in which contact line dissipation dominates the balance (e.g. when $b \gg \ell$ and $\zeta \gtrsim \mu$):
\begin{equation}
\rho g \sin{(\alpha)} \, UR^3 \sim \ell \zeta U^2\left(1+\frac{R}{b}\right)^{-2}
\end{equation}
resulting in
\begin{equation}
U \sim \frac{\gamma}{\zeta} \frac{R^3}{b^2 \kappa^{-1}} \left(1+b/R\right)^2 \sin{(\alpha)}.
\label{contact}
\end{equation}
In this case, the steady-state velocity is inversely proportional to the capillary length, so scales as $U \sim \sqrt{g}$, which is consistent with the behaviour observed in figure~\ref{FigVelg}. In the low slip limit ($b \ll R$), the velocity is proportional to the droplet volume while in the high slip limit ($b \gg R$) it is proportional to droplet radius (as seen in figure~\ref{FigVelR} for droplets with radii below the capillary length). 

It is evident that the scaling of $U$ observed in our molecular dynamics simulations is best explained by contact line dissipation in the high slip limit. In this limit, we expect $U \sim \frac{\gamma}{\zeta} \frac{R}{\kappa^{-1}} \sin{(\alpha)} + O\left(R/b\right)$. Of the dissipation mechanisms explored in this section, only this expression is consistent with both an increase in droplet speed with radius and an increase in speed with the strength of the gravitational body force that was observed in the simulations. This is also consistent with the decrease in $U$ observed as $\epsilon_{SL}$ increases (figure~\ref{FigEps}), as increasing the strength of the solid-liquid interaction will also cause $\zeta$ to increase.

\begin{figure}[htb]
\center
\resizebox{\columnwidth}{!}{\includegraphics{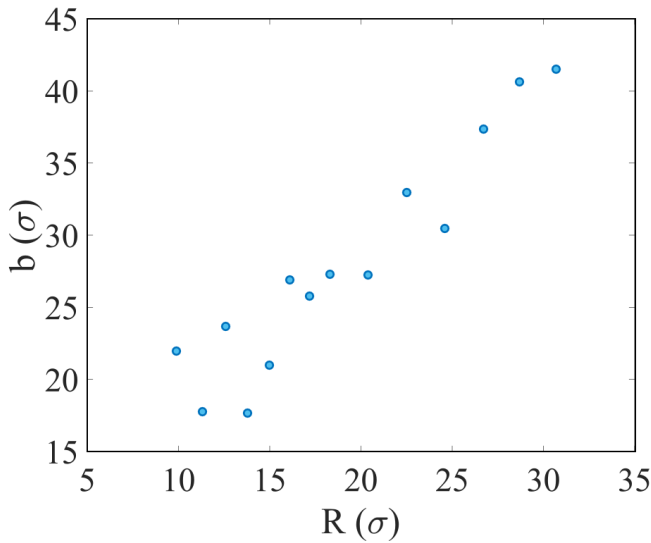}}
{\caption{The figure shows the estimated effective slip length $b \sim \frac{RU_s}{U_r}$ increasing with increasing droplet radius calculated for the $U_s$ and $U_r$ data shown in fig. \ref{FigVelR} (on a ridged surface, with $g=1\times 10^{-3}\sigma\tau^{-2}$ and $\epsilon_{SL}=0.125\epsilon$). We interpret the increase in {\em effective} slip length with $\epsilon$ as a decreasing contribution of the contact-line friction as droplet size increases.
\label{FigBvsR}}}
\end{figure}

Finally, this interpretation of the scaling of $U$ is consistent with our estimates of the magnitude of the slip length in the corresponding simulations. Calculating the slip length from the simulations is not straightforward (while $U_s$ is accessible, reliably estimating the shear at the surface is difficult), but we note that we can estimate $b$ from the approximation $U_s = b|\nabla u| \sim b (U_r/R) $ i.e. $b \sim R U_s/U_r$. This suggests that $b \sim 20 - 40 \sigma \gtrsim R$, as shown in figure~\ref{FigBvsR}, indicating that the droplets are experiencing high slip. A striking feature of figure~\ref{FigBvsR} is the increase in effective slip length $b$ with droplet size. We attribute this to the increasing importance of contact line friction as droplets get smaller, which is not explicitly captured in the Navier slip boundary condition but will likely enter as an effective contribution to the interfacial friction that scales as the ratio of contact zone perimeter to contact zone area.  

\section{Discussion}
It is interesting to consider the experimental application of the models developed in the previous section. Using molecular kinetic theory to describe the contact-line dynamics \cite{Blake69,Blake06}, an indicative value of the contact line friction parameter $\zeta$ for water on polyethelene terephthalate (contact angle 82$^{\circ}$) is 0.01 Pa$\cdot$s \cite{Blake06}, giving $\zeta / \mu \sim O(10)$. However, in molecular kinetic theory, $\zeta$ depends strongly on the contact angle $\theta_c$ \cite{Blake02}:
\begin{equation}
\zeta = \zeta_0 \exp \left( \frac{\gamma \lambda^2 \cos{(\theta_c)}}{k_B T}\right), 
\end{equation}
where $\zeta_0$ is the value of the friction coefficient at 90$^{\circ}$ and $\lambda$ is a molecular ``hopping" distance. Taking $\zeta_0 =$ 0.01 Pa$\cdot$s and $\lambda = 0.3$ nm, we estimate $\zeta / \mu \sim O(1)$ for water with a contact angle of 180$^{\circ}$ at room temperature.

For water, this suggests that $\zeta \sim \mu$ so the motion is unlikely to be dominated by interfacial friction even in the limit of large slip (as noted earlier, effective slip lengths on highly engineered superhydrophobic have been measured to be as large as 200 $\mu$m \cite{Kim08}). Thus for water on a superhydrophobic surface, the steady-state motion will either be dominated by viscous shear (when $b \ll \ell$), by contact line dissipation (when $b \gg \ell$) or all three forms of dissipation will be important (when $b \sim \ell$). It is possible that on close to ideal superhydrophobic surfaces, such as SLIPS surfaces \cite{wong2011bioinspired}, interfacial friction may become more important than contact line dissipation.  

\section{Conclusion}

In conclusion, we have extended a prior theoretical treatment of steady-state droplet motion on tilted superhydrophobic surfaces \cite{Mahadevan99} to account for interfacial and contact-line friction. We have investigated limiting cases where droplet motion is dominated by viscous dissipation, interfacial friction, or contact line dissipation in turn. Molecular dynamics simulations of droplets on tilted superhydrophobic surfaces reveal a mixture of slipping and rolling motion. The scaling behaviour of the simulated droplets with radii less than the capillary length suggest that the steady-state motion is dominated by contact line dissipation. On leaves, droplet motion is likely to be dominated by viscous dissipation, and slip can likely be neglected. Contact line dissipation, however, is likely to be important for highly engineered surfaces with high slip lengths, while interfacial friction may be important on close to ideal surfaces that lack features for contact line pinning to occur.

\begin{acknowledgments}
The authors thank the MacDiarmid Institute for Advanced Materials and Nanotechnology for funding. The authors also wish to acknowledge the contribution of NeSI high-performance computing facilities to the results of this research. New Zealand's national facilities are provided by the New Zealand eScience Infrastructure and funded jointly by NeSI's collaborator institutions and through the Ministry of Business, Innovation \& Employment's Research Infrastructure programme. URL https://www.nesi.org.nz.
\end{acknowledgments}

\bibliography{rolling_droplets}

\begin{thebibliography}{37}
\expandafter\ifx\csname natexlab\endcsname\relax\def\natexlab#1{#1}\fi
\expandafter\ifx\csname bibnamefont\endcsname\relax
  \def\bibnamefont#1{#1}\fi
\expandafter\ifx\csname bibfnamefont\endcsname\relax
  \def\bibfnamefont#1{#1}\fi
\expandafter\ifx\csname citenamefont\endcsname\relax
  \def\citenamefont#1{#1}\fi
\expandafter\ifx\csname url\endcsname\relax
  \def\url#1{\texttt{#1}}\fi
\expandafter\ifx\csname urlprefix\endcsname\relax\def\urlprefix{URL }\fi
\providecommand{\bibinfo}[2]{#2}
\providecommand{\eprint}[2][]{\url{#2}}

\bibitem[{\citenamefont{Barthlott and Neinhuis}(1997)}]{Barthlott97}
\bibinfo{author}{\bibfnamefont{W.}~\bibnamefont{Barthlott}} \bibnamefont{and}
  \bibinfo{author}{\bibfnamefont{C.}~\bibnamefont{Neinhuis}},
  \bibinfo{journal}{Planta} \textbf{\bibinfo{volume}{202}}, \bibinfo{pages}{1}
  (\bibinfo{year}{1997}).

\bibitem[{\citenamefont{Neinhuis and Barthlott}(1997)}]{Barthlott97b}
\bibinfo{author}{\bibfnamefont{C.}~\bibnamefont{Neinhuis}} \bibnamefont{and}
  \bibinfo{author}{\bibfnamefont{W.}~\bibnamefont{Barthlott}},
  \bibinfo{journal}{Ann Bot} \textbf{\bibinfo{volume}{79}},
  \bibinfo{pages}{667} (\bibinfo{year}{1997}).

\bibitem[{\citenamefont{Lafuma and Quere}(2003)}]{Quere03}
\bibinfo{author}{\bibfnamefont{A.}~\bibnamefont{Lafuma}} \bibnamefont{and}
  \bibinfo{author}{\bibfnamefont{D.}~\bibnamefont{Quere}},
  \bibinfo{journal}{Nat Mater} \textbf{\bibinfo{volume}{2}},
  \bibinfo{pages}{457} (\bibinfo{year}{2003}).

\bibitem[{\citenamefont{Fritsch et~al.}(2013)\citenamefont{Fritsch, Willmott,
  and Taylor}}]{fritsch2013superhydrophobic}
\bibinfo{author}{\bibfnamefont{A.}~\bibnamefont{Fritsch}},
  \bibinfo{author}{\bibfnamefont{G.}~\bibnamefont{Willmott}}, \bibnamefont{and}
  \bibinfo{author}{\bibfnamefont{M.}~\bibnamefont{Taylor}},
  \bibinfo{journal}{Journal of the Royal Society of New Zealand}
  \textbf{\bibinfo{volume}{43}}, \bibinfo{pages}{198} (\bibinfo{year}{2013}).

\bibitem[{\citenamefont{Quere}(2005)}]{Quere05}
\bibinfo{author}{\bibfnamefont{D.}~\bibnamefont{Quere}},
  \bibinfo{journal}{Reports on Progress in Physics}
  \textbf{\bibinfo{volume}{68}}, \bibinfo{pages}{2495} (\bibinfo{year}{2005}).

\bibitem[{\citenamefont{Richard and Quere}(1999)}]{Quere99}
\bibinfo{author}{\bibfnamefont{D.}~\bibnamefont{Richard}} \bibnamefont{and}
  \bibinfo{author}{\bibfnamefont{D.}~\bibnamefont{Quere}},
  \bibinfo{journal}{EPL (Europhysics Letters)} \textbf{\bibinfo{volume}{48}},
  \bibinfo{pages}{286} (\bibinfo{year}{1999}).

\bibitem[{\citenamefont{Cottin-Bizonne
  et~al.}(2003)\citenamefont{Cottin-Bizonne, Barrat, Bocquet, and
  Charlaix}}]{Cecile03}
\bibinfo{author}{\bibfnamefont{C.}~\bibnamefont{Cottin-Bizonne}},
  \bibinfo{author}{\bibfnamefont{J.-L.} \bibnamefont{Barrat}},
  \bibinfo{author}{\bibfnamefont{L.}~\bibnamefont{Bocquet}}, \bibnamefont{and}
  \bibinfo{author}{\bibfnamefont{E.}~\bibnamefont{Charlaix}},
  \bibinfo{journal}{Nat Mater} \textbf{\bibinfo{volume}{2}},
  \bibinfo{pages}{237} (\bibinfo{year}{2003}).

\bibitem[{\citenamefont{Rothstein}(2010)}]{Rothstein2010a}
\bibinfo{author}{\bibfnamefont{J.~P.} \bibnamefont{Rothstein}},
  \bibinfo{journal}{Annu. Rev. Fluid Mech} \textbf{\bibinfo{volume}{42}},
  \bibinfo{pages}{89} (\bibinfo{year}{2010}), ISSN \bibinfo{issn}{0066-4189}.

\bibitem[{\citenamefont{Choi and Kim}(2006)}]{Choi06}
\bibinfo{author}{\bibfnamefont{C.~H.} \bibnamefont{Choi}} \bibnamefont{and}
  \bibinfo{author}{\bibfnamefont{C.~J.} \bibnamefont{Kim}},
  \bibinfo{journal}{Physical Review Letters} \textbf{\bibinfo{volume}{96}},
  \bibinfo{pages}{066001} (\bibinfo{year}{2006}).

\bibitem[{\citenamefont{Joseph et~al.}(2006)\citenamefont{Joseph,
  Cottin-Bizonne, Beno\^{i}t, Ybert, Journet, Tabeling, and
  Bocquet}}]{Cecile06}
\bibinfo{author}{\bibfnamefont{P.}~\bibnamefont{Joseph}},
  \bibinfo{author}{\bibfnamefont{C.}~\bibnamefont{Cottin-Bizonne}},
  \bibinfo{author}{\bibfnamefont{J.~M.} \bibnamefont{Beno\^{i}t}},
  \bibinfo{author}{\bibfnamefont{C.}~\bibnamefont{Ybert}},
  \bibinfo{author}{\bibfnamefont{C.}~\bibnamefont{Journet}},
  \bibinfo{author}{\bibfnamefont{P.}~\bibnamefont{Tabeling}}, \bibnamefont{and}
  \bibinfo{author}{\bibfnamefont{L.}~\bibnamefont{Bocquet}},
  \bibinfo{journal}{Physical Review Letters} \textbf{\bibinfo{volume}{97}},
  \bibinfo{pages}{156104} (\bibinfo{year}{2006}).

\bibitem[{\citenamefont{Ybert et~al.}(2007)\citenamefont{Ybert, Baretin,
  Bizonne, Joseph, and Bocquet}}]{Ybert07}
\bibinfo{author}{\bibfnamefont{C.}~\bibnamefont{Ybert}},
  \bibinfo{author}{\bibfnamefont{C.}~\bibnamefont{Baretin}},
  \bibinfo{author}{\bibfnamefont{C.~C.} \bibnamefont{Bizonne}},
  \bibinfo{author}{\bibfnamefont{P.}~\bibnamefont{Joseph}}, \bibnamefont{and}
  \bibinfo{author}{\bibfnamefont{L.}~\bibnamefont{Bocquet}},
  \bibinfo{journal}{Phys. Fluids} \textbf{\bibinfo{volume}{19}},
  \bibinfo{pages}{123601} (\bibinfo{year}{2007}).

\bibitem[{\citenamefont{Lee et~al.}(2008)\citenamefont{Lee, Choi, and
  Kim}}]{Kim08}
\bibinfo{author}{\bibfnamefont{C.}~\bibnamefont{Lee}},
  \bibinfo{author}{\bibfnamefont{C.~H.} \bibnamefont{Choi}}, \bibnamefont{and}
  \bibinfo{author}{\bibfnamefont{C.~J.~C.} \bibnamefont{Kim}},
  \bibinfo{journal}{Physical Review Letters} \textbf{\bibinfo{volume}{101}},
  \bibinfo{pages}{064501+} (\bibinfo{year}{2008}).

\bibitem[{\citenamefont{Mahadevan and Pomeau}(1999)}]{Mahadevan99}
\bibinfo{author}{\bibfnamefont{L.}~\bibnamefont{Mahadevan}} \bibnamefont{and}
  \bibinfo{author}{\bibfnamefont{Y.}~\bibnamefont{Pomeau}},
  \bibinfo{journal}{Physics of Fluids} \textbf{\bibinfo{volume}{11}},
  \bibinfo{pages}{2449} (\bibinfo{year}{1999}).

\bibitem[{\citenamefont{Sakai et~al.}(2006)\citenamefont{Sakai, Song, Yoshida,
  Suzuki, Kameshima, and Nakajima}}]{sakai06}
\bibinfo{author}{\bibfnamefont{M.}~\bibnamefont{Sakai}},
  \bibinfo{author}{\bibfnamefont{J.-H.} \bibnamefont{Song}},
  \bibinfo{author}{\bibfnamefont{N.}~\bibnamefont{Yoshida}},
  \bibinfo{author}{\bibfnamefont{S.}~\bibnamefont{Suzuki}},
  \bibinfo{author}{\bibfnamefont{Y.}~\bibnamefont{Kameshima}},
  \bibnamefont{and} \bibinfo{author}{\bibfnamefont{A.}~\bibnamefont{Nakajima}},
  \bibinfo{journal}{Langmuir} \textbf{\bibinfo{volume}{22}},
  \bibinfo{pages}{4906} (\bibinfo{year}{2006}).

\bibitem[{\citenamefont{Sakai et~al.}(0000)\citenamefont{Sakai, Kono, Nakajima,
  Zhang, Sakai, Abe, and Fujishima}}]{sakai09}
\bibinfo{author}{\bibfnamefont{M.}~\bibnamefont{Sakai}},
  \bibinfo{author}{\bibfnamefont{H.}~\bibnamefont{Kono}},
  \bibinfo{author}{\bibfnamefont{A.}~\bibnamefont{Nakajima}},
  \bibinfo{author}{\bibfnamefont{X.}~\bibnamefont{Zhang}},
  \bibinfo{author}{\bibfnamefont{H.}~\bibnamefont{Sakai}},
  \bibinfo{author}{\bibfnamefont{M.}~\bibnamefont{Abe}}, \bibnamefont{and}
  \bibinfo{author}{\bibfnamefont{A.}~\bibnamefont{Fujishima}},
  \bibinfo{journal}{Langmuir} \textbf{\bibinfo{volume}{0}}
  (\bibinfo{year}{0000}).

\bibitem[{\citenamefont{Blake and Haynes}(1969)}]{Blake69}
\bibinfo{author}{\bibfnamefont{T.~D.} \bibnamefont{Blake}} \bibnamefont{and}
  \bibinfo{author}{\bibfnamefont{J.~M.} \bibnamefont{Haynes}},
  \bibinfo{journal}{Journal of Colloid and Interface Science}
  \textbf{\bibinfo{volume}{30}}, \bibinfo{pages}{421 } (\bibinfo{year}{1969}).

\bibitem[{\citenamefont{Blake and Coninck}(2002)}]{Blake02}
\bibinfo{author}{\bibfnamefont{T.~D.} \bibnamefont{Blake}} \bibnamefont{and}
  \bibinfo{author}{\bibfnamefont{J.~D.} \bibnamefont{Coninck}},
  \bibinfo{journal}{Advances in Colloid and Interface Science}
  \textbf{\bibinfo{volume}{96}}, \bibinfo{pages}{21 } (\bibinfo{year}{2002}).

\bibitem[{\citenamefont{Blake}(2006)}]{Blake06}
\bibinfo{author}{\bibfnamefont{T.~D.} \bibnamefont{Blake}},
  \bibinfo{journal}{Journal of Colloid and Interface Science}
  \textbf{\bibinfo{volume}{299}}, \bibinfo{pages}{1 } (\bibinfo{year}{2006}).

\bibitem[{\citenamefont{Olin et~al.}(2013)\citenamefont{Olin, Lindstr{\"{o}}m,
  Pettersson, and W{\aa}gberg}}]{Olin2013}
\bibinfo{author}{\bibfnamefont{P.}~\bibnamefont{Olin}},
  \bibinfo{author}{\bibfnamefont{S.~B.} \bibnamefont{Lindstr{\"{o}}m}},
  \bibinfo{author}{\bibfnamefont{T.}~\bibnamefont{Pettersson}},
  \bibnamefont{and}
  \bibinfo{author}{\bibfnamefont{L.}~\bibnamefont{W{\aa}gberg}},
  \bibinfo{journal}{Langmuir} \textbf{\bibinfo{volume}{29}},
  \bibinfo{pages}{9079} (\bibinfo{year}{2013}), ISSN \bibinfo{issn}{07437463}.

\bibitem[{\citenamefont{Butt et~al.}(2017)\citenamefont{Butt, Gao,
  Papadopoulos, Steffen, Kappl, and {Diger Berger}}}]{Butt2017}
\bibinfo{author}{\bibfnamefont{H.-J.} \bibnamefont{Butt}},
  \bibinfo{author}{\bibfnamefont{N.}~\bibnamefont{Gao}},
  \bibinfo{author}{\bibfnamefont{P.}~\bibnamefont{Papadopoulos}},
  \bibinfo{author}{\bibfnamefont{W.}~\bibnamefont{Steffen}},
  \bibinfo{author}{\bibfnamefont{M.}~\bibnamefont{Kappl}}, \bibnamefont{and}
  \bibinfo{author}{\bibfnamefont{R.}~\bibnamefont{{Diger Berger}}}
  (\bibinfo{year}{2017}), ISSN \bibinfo{issn}{0743-7463}.

\bibitem[{\citenamefont{Servantie and Moller}(2008)}]{Servantie2008}
\bibinfo{author}{\bibfnamefont{J.}~\bibnamefont{Servantie}} \bibnamefont{and}
  \bibinfo{author}{\bibfnamefont{M.}~\bibnamefont{Moller}},
  \bibinfo{journal}{Journal of Chemical Physics} \textbf{\bibinfo{volume}{128}}
  (\bibinfo{year}{2008}), ISSN \bibinfo{issn}{00219606}, \eprint{0708.1944}.

\bibitem[{\citenamefont{Annapragada et~al.}(2012)\citenamefont{Annapragada,
  Murthy, and Garimella}}]{Annapragada2012}
\bibinfo{author}{\bibfnamefont{S.~R.} \bibnamefont{Annapragada}},
  \bibinfo{author}{\bibfnamefont{J.~Y.} \bibnamefont{Murthy}},
  \bibnamefont{and} \bibinfo{author}{\bibfnamefont{S.~V.}
  \bibnamefont{Garimella}}, \bibinfo{journal}{International Journal of Heat and
  Mass Transfer} \textbf{\bibinfo{volume}{55}}, \bibinfo{pages}{1457}
  (\bibinfo{year}{2012}).

\bibitem[{\citenamefont{Mognetti et~al.}(2010)\citenamefont{Mognetti,
  Kusumaatmaja, and Yeomans}}]{Mognetti2010}
\bibinfo{author}{\bibfnamefont{B.~M.} \bibnamefont{Mognetti}},
  \bibinfo{author}{\bibfnamefont{H.}~\bibnamefont{Kusumaatmaja}},
  \bibnamefont{and} \bibinfo{author}{\bibfnamefont{J.~M.}
  \bibnamefont{Yeomans}}, \bibinfo{journal}{Faraday discussions}
  \textbf{\bibinfo{volume}{146}}, \bibinfo{pages}{153} (\bibinfo{year}{2010}),
  ISSN \bibinfo{issn}{1359-6640}, \eprint{1009.4658},
  \urlprefix\url{http://www.ncbi.nlm.nih.gov/pubmed/21043420}.

\bibitem[{\citenamefont{Li et~al.}(2013)\citenamefont{Li, Hu, Wang, Ma, and
  Zhou}}]{Li2013}
\bibinfo{author}{\bibfnamefont{Z.}~\bibnamefont{Li}},
  \bibinfo{author}{\bibfnamefont{G.~H.} \bibnamefont{Hu}},
  \bibinfo{author}{\bibfnamefont{Z.~L.} \bibnamefont{Wang}},
  \bibinfo{author}{\bibfnamefont{Y.~B.} \bibnamefont{Ma}}, \bibnamefont{and}
  \bibinfo{author}{\bibfnamefont{Z.~W.} \bibnamefont{Zhou}},
  \bibinfo{journal}{Physics of Fluids} \textbf{\bibinfo{volume}{25}}
  (\bibinfo{year}{2013}), ISSN \bibinfo{issn}{10706631}.

\bibitem[{\citenamefont{Wind-Willassen and
  S{\o}rensen}(2014)}]{Wind-Willassen2014}
\bibinfo{author}{\bibfnamefont{{\O}.}~\bibnamefont{Wind-Willassen}}
  \bibnamefont{and} \bibinfo{author}{\bibfnamefont{M.~P.}
  \bibnamefont{S{\o}rensen}}, \bibinfo{journal}{European Physical Journal E}
  \textbf{\bibinfo{volume}{37}}, \bibinfo{pages}{1} (\bibinfo{year}{2014}),
  ISSN \bibinfo{issn}{1292895X}.

\bibitem[{\citenamefont{Karapetsas et~al.}(2016)\citenamefont{Karapetsas,
  Chamakos, and Papathanasiou}}]{Karapetsas2016}
\bibinfo{author}{\bibfnamefont{G.}~\bibnamefont{Karapetsas}},
  \bibinfo{author}{\bibfnamefont{N.~T.} \bibnamefont{Chamakos}},
  \bibnamefont{and} \bibinfo{author}{\bibfnamefont{A.~G.}
  \bibnamefont{Papathanasiou}}, \bibinfo{journal}{Journal of Physics: Condensed
  Matter} \textbf{\bibinfo{volume}{28}}, \bibinfo{pages}{085101}
  (\bibinfo{year}{2016}), ISSN \bibinfo{issn}{0953-8984},
  \urlprefix\url{http://stacks.iop.org/0953-8984/28/i=8/a=085101?key=crossref.d19b3c81d5669674f52d55565f04aec2}.

\bibitem[{\citenamefont{Hendy and Lund}(2007)}]{Hendy07}
\bibinfo{author}{\bibfnamefont{S.~C.} \bibnamefont{Hendy}} \bibnamefont{and}
  \bibinfo{author}{\bibfnamefont{N.~J.} \bibnamefont{Lund}},
  \bibinfo{journal}{Physical Review E (Statistical, Nonlinear, and Soft Matter
  Physics)} \textbf{\bibinfo{volume}{76}}, \bibinfo{pages}{066313}
  (\bibinfo{year}{2007}).

\bibitem[{\citenamefont{Plimpton}(1995)}]{LAMMPS}
\bibinfo{author}{\bibfnamefont{S.~J.} \bibnamefont{Plimpton}},
  \bibinfo{journal}{Journal of Computational Physics}
  \textbf{\bibinfo{volume}{117}}, \bibinfo{pages}{1} (\bibinfo{year}{1995}).

\bibitem[{\citenamefont{Yong and Zhang}(2009)}]{Yong2009}
\bibinfo{author}{\bibfnamefont{X.}~\bibnamefont{Yong}} \bibnamefont{and}
  \bibinfo{author}{\bibfnamefont{L.~T.} \bibnamefont{Zhang}},
  \bibinfo{journal}{Langmuir} \textbf{\bibinfo{volume}{25}},
  \bibinfo{pages}{5045} (\bibinfo{year}{2009}), ISSN \bibinfo{issn}{07437463}.

\bibitem[{\citenamefont{Schneider and Stoll}(1978)}]{Schneider78}
\bibinfo{author}{\bibfnamefont{T.}~\bibnamefont{Schneider}} \bibnamefont{and}
  \bibinfo{author}{\bibfnamefont{E.}~\bibnamefont{Stoll}},
  \bibinfo{journal}{Phys. Rev. B} \textbf{\bibinfo{volume}{17}},
  \bibinfo{pages}{1302} (\bibinfo{year}{1978}),
  \urlprefix\url{http://link.aps.org/doi/10.1103/PhysRevB.17.1302}.

\bibitem[{\citenamefont{Hess}(2002)}]{Hess:2002p8998}
\bibinfo{author}{\bibfnamefont{B.}~\bibnamefont{Hess}}, \bibinfo{journal}{The
  Journal of Chemical Physics} \textbf{\bibinfo{volume}{116}},
  \bibinfo{pages}{209} (\bibinfo{year}{2002}).

\bibitem[{\citenamefont{Rowley and Painter}(1997)}]{Rowley:1997p8980}
\bibinfo{author}{\bibfnamefont{R.}~\bibnamefont{Rowley}} \bibnamefont{and}
  \bibinfo{author}{\bibfnamefont{M.}~\bibnamefont{Painter}},
  \bibinfo{journal}{International Journal of Thermophysics}
  \textbf{\bibinfo{volume}{18}}, \bibinfo{pages}{1109} (\bibinfo{year}{1997}).

\bibitem[{\citenamefont{Tankeshwar et~al.}(1988)\citenamefont{Tankeshwar,
  Pathak, and Ranganathan}}]{Tankeshwar:1988p8991}
\bibinfo{author}{\bibfnamefont{K.}~\bibnamefont{Tankeshwar}},
  \bibinfo{author}{\bibfnamefont{K.}~\bibnamefont{Pathak}}, \bibnamefont{and}
  \bibinfo{author}{\bibfnamefont{S.}~\bibnamefont{Ranganathan}},
  \bibinfo{journal}{Journal of Physics C: Solid State Physics}
  \textbf{\bibinfo{volume}{21}}, \bibinfo{pages}{3607} (\bibinfo{year}{1988}).

\bibitem[{\citenamefont{Lee and Stein}(1987)}]{Lee:1987p8281}
\bibinfo{author}{\bibfnamefont{J.}~\bibnamefont{Lee}} \bibnamefont{and}
  \bibinfo{author}{\bibfnamefont{G.}~\bibnamefont{Stein}},
  \bibinfo{journal}{Journal of Physical Chemistry}  (\bibinfo{year}{1987}).

\bibitem[{\citenamefont{Potoff and Panagiotopoulos}(2000)}]{Potoff:2000p8106}
\bibinfo{author}{\bibfnamefont{J.}~\bibnamefont{Potoff}} \bibnamefont{and}
  \bibinfo{author}{\bibfnamefont{A.}~\bibnamefont{Panagiotopoulos}},
  \bibinfo{journal}{The Journal of Chemical Physics}  (\bibinfo{year}{2000}).

\bibitem[{\citenamefont{Brochard-Wyart and de~Gennes}(1992)}]{deGennes92}
\bibinfo{author}{\bibfnamefont{F.}~\bibnamefont{Brochard-Wyart}}
  \bibnamefont{and}
  \bibinfo{author}{\bibfnamefont{P.}~\bibnamefont{de~Gennes}},
  \bibinfo{journal}{Advances in Colloid and Interface Science}
  \textbf{\bibinfo{volume}{39}}, \bibinfo{pages}{1} (\bibinfo{year}{1992}).

\bibitem[{\citenamefont{Wong et~al.}(2011)\citenamefont{Wong, Kang, Tang,
  Smythe, Hatton, Grinthal, and Aizenberg}}]{wong2011bioinspired}
\bibinfo{author}{\bibfnamefont{T.-S.} \bibnamefont{Wong}},
  \bibinfo{author}{\bibfnamefont{S.~H.} \bibnamefont{Kang}},
  \bibinfo{author}{\bibfnamefont{S.~K.} \bibnamefont{Tang}},
  \bibinfo{author}{\bibfnamefont{E.~J.} \bibnamefont{Smythe}},
  \bibinfo{author}{\bibfnamefont{B.~D.} \bibnamefont{Hatton}},
  \bibinfo{author}{\bibfnamefont{A.}~\bibnamefont{Grinthal}}, \bibnamefont{and}
  \bibinfo{author}{\bibfnamefont{J.}~\bibnamefont{Aizenberg}},
  \bibinfo{journal}{Nature} \textbf{\bibinfo{volume}{477}},
  \bibinfo{pages}{443} (\bibinfo{year}{2011}).

\end{thebibliography}

\end{document}